\documentstyle[12pt,russian]{article}
\begin{document}
\setcounter{page}{160}
\fussy
\flushbottom
\parindent=0.25in
\oddsidemargin=0.75in
\evensidemargin=0.75in
\topmargin=1in
\headheight=0.1in
\headsep=0.2in
\footskip=0.3in
\footheight=0.3in
\textheight=6.7in
\textwidth=4.7in
\noindent 
\bigskip

{\small
\begin{center} {\bf GEOMETRICAL INTERPRETATION OF 
QUANTUM FORMALISM}
\end{center}
\smallskip
\begin{center}O. A. OLKHOV\\
{\it N. Semenov Institute of Chemical Physics, Russian Academy of Sciences, 
Kosygin Street 4, Moscow, 117977 Russia\\
E-mail: olkhov@center.chph.ras.ru}\end{center}
\noindent
\smallskip
Proceedings of the 7th International Symposium on Particles, Strings and
Cosmology (PASCOS 99), Lake Tahoe, California, 10-16 Desember 1999;
Singapure-New Jersey-Hong Kong.: World Scientific, 2000, P.160
\smallskip

\noindent {\footnotesize We show that the Dirac equation can be
rewritten as a relation describing the fundamental symmetry group of special
topological manifold corresponding to the Dirac wave field. It leads to
unification of the time-space and internal symmetries wihin one symmetry group.
We suppose
that nonelectromagnetic interactions appear within such approach as the
deviations of the metrics from the euclidean form. The
expression for the long-range part of "effective" nucleon-nucleon potential
is derived using this assumption.}
\par \bigskip

In this papaer we attempt to introduce the topological and  
group-theoretical approaches
at the very beginning of quantum formalism. Let us show,
at first, that the equation for the free Dirac field can be
considered as a relation describing 
symmetry proprieties of the nonorientable local Euclidean Riemann 
time-space representing this Dirac wave field. 

For free field the Dirac equation can be written as [1]:
$$\left(i\frac {\partial}{\partial x_0}\gamma_0-i\frac {\partial}{\partial x_1}
\gamma_1-i\frac {\partial}{\partial x_2}\gamma_2-i\frac {\partial}
{\partial x_3}\gamma_3\right)\psi=K_M\psi, \eqno (1)$$
where $x_0=ct,\quad x_1=x,\quad x_2=y,\quad x_3=z$ and $K_M=mc/\hbar.$ 
In (1) $\psi$ is
the four-component spinor and $\gamma_l$ are the known 
four-dimensional matrices satisfying the following relationships:
$$\gamma_l\gamma_n+\gamma_n\gamma_l=0,\quad l\ne n;\quad \gamma_0^2=1,\quad
\gamma_1^2=\gamma_2^2=\gamma_3^2=-1.\eqno (2)$$
We write the energy conservation low $E^2-p^2c^2=m^2c^4$ in the form
$$K_0^2-K_1^2-K_2^2-K_3^2=K_M^2,\eqno (3)$$
where $K_0=E/\hbar c$ and $K_{1,2,3}=p_{1,2,3}/\hbar$.

We shall show now that operators $K_n^{-1}i\partial/\partial x_n$ and matrices
$\gamma_n$ in (1) can be considered as the representations 
of two time-space symmetry transformations: translations and reflections.

Let us show that $\hat T=K_n^{-1}i\partial/\partial x_n$ is the operator of
translation at distance $\Delta X_n=2\pi/K_n$. The $x_n$-dependence of 
function $\psi$ in (1) has the form $\psi \sim exp(-iK_nx_n)$ [1]. Hence,
$$i\partial/\partial x_n\psi=K_nexp(-iK_nx_n)=
K_nexp[-iK_n(x_n+2\pi/K_n)]=K_n\psi (x_n+\Delta X_n).$$
This can be rewritten as
$$\hat T\psi(x_n)=\psi (x_n+\Delta X_n),\eqno (4)$$
which signifies that $\hat T(\Delta X_n)$ is, in fact, the operator 
of translation at distance $\Delta X_n$ [2].

Let now show that $\gamma_n$ are the reflection operators $\hat R_n$: 
$R_1\psi (x_0,x_1,x_2,x_3)=\psi'(-x_0,x_1,-x_2,-x_3)$ etc.
As to the matrix $\gamma_0$, it is
known to be the reflection matrix $\hat R_0$ (the 
so called "space reflection" [1]). Let us consider, for example, the
product $\gamma_2\gamma_3$. This product formally represents the spatial 
rotation 
about the $x_1$-axis by the angle $\pi$ [1]. 
At the same time, the same rotation can be accomplished by the two successive
reflections: ($x_2'=x_2, x_{0,1,3}'=
-x_{0,1,3}$) and ($x_3'=x_3, x_{0,1,2}'=-x_{0,1,2}$). Hence, $\gamma_2$ 
and $\gamma_3$ are, in fact, the $\hat R_2$ and $\hat R_3$ reflections.

Finally we can rewrite Eq.(1) in the following form:
$$(K_0\hat T_0\hat R_0-K_1\hat T_1\hat R_1-K_2\hat T_2\hat R_2
-K_3\hat T_3\hat R_3)\psi=K_M\psi.\eqno (5)$$

What does it mean?  We suggest a new interpretation for this equation.
Let us consider this equation as a 
group-theoretical relation describing the geometrical proprieties of a certain
distorted Minkovski space, a manifold that represents Dirac free wave field.
Within such an approach, the pseudoeuclidean Minkovski space plays the role of
a universal covering surface for the manifold, while the $\psi$-function 
becomes the basis vector of a representation of the corresponding fundamental
group.

Spinor representation in Eq.(5) implies that this equation discribes
symmetry properties of a certain kind of spinor geometric object. 
Such an
object has rather specific proprieties typical of a nonorientable local
Euclidean Riemann space. For example, it reverts to 
the initial position not after
rotation by angle $2\pi$ but after rotation by angle $4\pi$ (as the M\"obius
strip) [3,4]. We see that the spinor geometrical object in Eq.(5) 
behaves in the same way after certain reflections
($\gamma_{1,2,3}^2=-1$, but $\gamma_{1,2,3}^4=+1$). 
It is also known that some nonorientable manifolds have a fundamental group
generated by glide reflections (product of translation and reflection [5]),
and we see that Eq.(5) contains operators of glide reflection.

As a result we can state that Dirac equation (1) is, in fact, the
group-theoretical relation describing symmetry properties of a special
quantum object, nonorientable local Euclidean Riemann space
representing the free Dirac wave field. 
Energy, momentum, mass, spin and charge 
become the corresponding topological invariants and the Dirac theory
appears as geometrical theory for the curved four-dimentional manifold. 

The above approach can easilly be generalized to the classical Maxwell wave
field and we'll discuss this case in details into subsequent publications.
(Unlike the Dirac field, this field appears as the orientable local Euclidean
space).
In any case, the usefullness of the new interpretation can be established
after its application to the problem of interacting fields.
Moving in this direction, we start with the simplest  
one-particle approximation when the Dirac
field can be considered as classical and the external field can be
considered as a given function. 

Let us consider the problem of electron in a
hydrodgen atom. Dirac equation (5) for this case has the form (with our
notations) [1]:
$$K_0(\hat T_0-\varphi (r))\hat R_0-K_1\hat T_1\hat R_1-K_2\hat T_2\hat R_2-
K_3\hat T_3\hat R_3)\psi =K_M\psi,$$
where $\varphi (r)$ is the Coulomb potential.                                                                                                                                                                                                                                                                                                                                                                                                                                                                               
We see that within the topological interpretation, "switching on" of 
the external electrostatic
field means the violation of translational
symmetry: this violation leads to an "inhomogeneous compression" of the 
initial manifold along the $x_0$-axis.

Assume that the nonelectromagnetic fields (nuclear ones, for example) also,
within one-particle approximation, violate the initial symmetry
of Riemann space generated by the free Dirac field. These violations
should be different from these caused by the static electromagnetic field 
(violation
of translational symmetry). They should be violations of the reflection
symmetry, the second type of initial symmetry. Such violations can distort the
initial local Eucledian metrics. In this case Eq.(5) can be
written as
$$\sum_{l,n=1}^4 g_{ln}(x)K_l\hat T_l\hat R_n\psi=K_M\psi,\eqno (6)$$
where the undistorted metrics $g_{ln}(x)$ had the pseudoeucledian form:
$g_{00}=1,g_{11}=g_{22}=g_{33}=-1, g_{ln}=0, l\ne n$.

To clarify whether the above hypothesis is realistic 
we used the following simplified form for $g_{ln}(x)$: $g_{01}=g_{10}=
g_{02}=g_{20}=g_{03}=g_{30}=0,\quad g_{00}=1-v_1(r),\quad g_{11}=g_{22}=
g_{33}=1-v_2(r), g_{12}=g_{13}=g_{23}=1-v_3(r)$,\quad where $r^2=x_1^2+
x_2^2+x_3^2$. In the nonrelativistic limit of Eq.(6), we obtain (using the
standard method [1]) Schr\"odinger equation for the stationary state 
with the "effective" potential
$$\hat V_{eff}(r,E,\hat P)=Ev_1(r)+\frac {1}{2m}\hat P^2v_2(r)+
\frac{1}{m}(\hat P_x\hat P_y+\hat P_x\hat P_y+\hat P_Y\hat P_z)v_3(r),
\eqno (7)$$
where $\hat P_{x,y,z}$ are the momentum operators, $E$ is energy and 
$v(r)$ are unknown functions. Interestingly, thus obtained
effective potential (7) is very close to the semiphenomenological so called
"Paris nucleon-nucleon 
potential", which was successfully used for the interpretation of
experimental data on nucleon-nucleon scattering [6].
\par
\bigskip
\noindent
{\bf References}
\par\smallskip    
1. Schweber S., {\it An introduction to relativistic quantum field theory} 
(Row, 

Peterson and Co., Evanstone, Ill., Elmsford, N.Y. 1961).

2. Heine V., {\it Group theory in quantum mechanics} (Pergamon Press, London---

N.Y.---Paris, 1960).

3. Cartan M.E., {\it Le\c{c}ons sur la th\'eorie des spineurs} 
(Sorbonne, 1938).

4. Penrouse R., Rindler W., {\it Spinors and space-time, V.1} (Cambridge 
University 

Press, Cambridge, 1984).

5. Coxeter H.S.M., {\it Introduction to geometry} (John Willey and Sons, Inc., 

N.Y.---London, 1961).

6. R. Vinh Mau, in {\it Mesons in Nuclei, V.1, P. 151}, 
ed. Mannque Rho, Denys 

Wilkinson (North---Holland Publ. Co., Amsterdam---N.Y.---Oxford, 1979)

\end{document}